\documentclass[runningheads]{llncs}
\usepackage[T1]{fontenc}
\usepackage{cite}
\usepackage{amsmath,amssymb,amsfonts}
\usepackage{algorithmic}
\usepackage{graphicx}
\usepackage{textcomp}
\usepackage{tikz}
\usepackage{tcolorbox}
\usepackage{todonotes}
\usepackage{adjustbox}
\usetikzlibrary{positioning}
\usetikzlibrary{tikzmark}
\usetikzlibrary{fit}
\usetikzlibrary{matrix}
\usepackage[ruled,vlined]{algorithm2e}
\usepackage{url}

\begin{document}

\title{From RDF Graph Validation to RDF Dataset Validation with SHACL-DS}

\author{
    Davan Chiem Dao\orcidID{0009-0004-8139-1927} \and \\
    Christophe Debruyne \orcidID{0000-0003-4734-3847}
}

\authorrunning{D. Chiem Dao et al.}

\institute{Montefiore Institute, University of Liège, Liège, Belgium\\
\email{\{davan.chiemdao,c.debruyne\}@uliege.be}}


\maketitle

\begin{abstract}
The Shapes Constraint Language (SHACL) is the W3C Recommendation for validating a single RDF graph. This makes SHACL inadequate for validating data across (named) graphs in an RDF dataset. Existing workarounds, such as graph unions or bespoke preprocessing, either collapse the RDF dataset structure or compromise the declarative nature of SHACL validation. In the former, we lose track of where triples come from; in the latter, knowledge is hidden in the code, and the constraints are not self-contained nor fully declarative. We present SHACL-DS to address this problem. SHACL-DS proposes a vocabulary and an algorithm on top of SHACL for RDF dataset validation. SHACL-DS introduces the concepts of Shapes Datasets, Target Graph Declarations, and Target Graph Combinations, enabling declarative constraints to operate across multiple graphs in an RDF dataset. SHACL-DS also defines the behaviour of SPARQL-based constraints for validating RDF datasets. In this paper, we formalize SHACL-DS and provide a prototype implementation.
\end{abstract}

\section{Introduction}
\label{sec:introduction}
SHACL \cite{shacl} is a W3C Recommendation that provides a powerful mechanism for validating RDF data against structural and semantic constraints. Those constraints are expressed declaratively, meaning that SHACL implementations can adopt different approaches to validate them as long as they adhere to the specification. In SHACL, one defines the expected structure (the "shape") that the data in an RDF graph should conform to, and the SHACL implementation then checks whether the data complies with those constraints. The validation process takes as input two RDF graphs: the shapes graph and the data graph. The first contains the SHACL shapes and constraints that define the expected structure of the latter. The output is a new RDF graph that describes the validation outcome. As IRIs are used to identify resources in the data graph, those IRIs would be used in the validation report to pinpoint where the problem lies.

SHACL validation is defined only for a single RDF graph. This assumption leads to problems when data is represented as datasets comprising multiple named graphs, for example, to model provenance or organize different subsets of a knowledge graph project, and one wants to validate specific (combinations of) graphs. When applying SHACL to RDF datasets, practitioners face a dilemma: either "flatten" the dataset into a single graph by taking the union and thereby losing the provenance of each triple, or implement bespoke logic to iterate over and validate (subsets of) named graphs. In other words, validation has become partly procedural, with knowledge (i.e., which graphs to validate together, and how) hidden in the bespoke code. Both approaches are error-prone and, in our opinion, conceptually unsound. 

This study aims to address the following research problem: \textit{"How can SHACL be extended to validate RDF datasets, and preserve backward compatibility with the original specification?"} Our answer to this problem is SHACL-DS, which builds upon SHACL to provide RDF dataset validation. SHACL-DS introduces a lightweight vocabulary and semantics that:

\begin{itemize}
    \item Associate named shapes graphs with specific named data graphs within an RDF dataset through explicit targeting declarations;
    \item Combine named graphs in RDF dataset into validation targets using set operations such as union, intersection, and difference; and
    \item Prescribes how SPARQL-based constraints should be evaluated against validation targets, allowing for reusable cross-graph rules.
\end{itemize}

With these, SHACL-DS can process and validate RDF datasets as first-class entities. We implemented a prototype by extending the SHACL module of dotNetRDF. The prototype also includes a test suite that covers all features introduced in this article. 

The remainder of this paper is organized as follows: Section 2 describes and motivates the problem, Section 3 presents related work and the state of the art, Section 4 introduces and formalizes our approach, called SHACL-DS, and Section 5 presents its implementation. In Sections 6 and 7, we respectively discuss our contributions and conclude our paper.

\section{Motivating Example and Problem}
We first identified the problem of validating RDF datasets in a KG project in the \cite{endorse2023paper} domain. In this project, data from multiple sources, updated quarterly, were stored in named graphs to track their provenance. Whilst SHACL allowed us to model most constraints, we faced issues when constraints depended on certain (combinations) of graphs. Because SHACL validates a single data graph, triples from multiple graphs must first be merged (“flattened”) into a single graph. This preprocessing step obscures provenance and can lead to validation bypasses \cite{shacl_bypass}: triples originating from one graph may affect constraints that should apply only when data comes from another graph. In other words, flattening erases contextual cues from named graphs, allowing erroneous or misplaced triples to pass validation undetected. Another approach is to stage graphs for different validation purposes. Still, it has the disadvantage that the RDF dataset validation process becomes less declarative and does not scale to complex constraints (e.g., an \texttt{sh:and}) across different graphs.

This observation motivated the development of SHACL-DS, an approach that builds on SHACL to enable validation at the dataset level, treating named graphs and their relationships as first-class citizens. By preserving provenance and supporting cross-graph validation, SHACL-DS addresses the limitations uncovered in this use case.

\section{Related Work and State of the Art}
SHACL has been investigated from multiple perspectives. Research on shape discovery and induction aims to automatically derive SHACL shapes from data \cite{rdfminer,learnSHACL,AutoExtractSHACL} or from existing mappings \cite{RML2SHACL}. Other studies explore theoretical foundations, positioning SHACL within formal frameworks such as Description Logics \cite{SHACL-DL,SHACL-DL2} and OWL \cite{shaclOwl}. On the practical side, SHACL has been applied to diverse tasks including access control \cite{SHACL-ACL} and form generation \cite{schimatos,shacl-form}. Yet, despite this breadth of work, the problem of validating RDF datasets comprising multiple named graphs has received little attention.

This section explores the state of the art in RDF dataset validation. We begin by reviewing the SHACL specification, focusing on elements that may be implicitly related to or interact with RDF datasets. The implicit and unclear nature is deemed \textit{problematic} in the context of RDF Dataset validation. Next, we review how some SHACL implementations handle the validation of RDF datasets, which will \textit{highlight current limitations}. Lastly, we take a closer look at \cite{shacl_x}, which \textit{proposed} a SHACL extension to RDF datasets. This proposal provided no algorithm and no prototype implementation to the best of our knowledge. It was thus a vocabulary extension, but it did offer a starting point for the development of SHACL-DS.

By analyzing these aspects, we aim to identify the gaps in the current state of the art and establish the context for developing SHACL-DS.

\subsection{SHACL}
In SHACL, constraints are grouped into a single RDF shapes graph, and these are applied to a single RDF data graph. This "single-graph" model is central to SHACL’s design and limits its direct applicability to RDF datasets. Nevertheless, several aspects of the SHACL and SHACL-SPARQL specifications implicitly suggest that SHACL could be extended toward some RDF dataset validation:

\begin{itemize}
    \item SHACL states that a shapes graph \textit{can} be a union of graphs, but the graphs themselves are then lost.
    \item A shapes graph containing \texttt{owl:imports} \textit{should} be extended with the RDF graphs referenced by this predicate. While not explicitly a "dataset-level"\footnote{With "dataset-level," we refer to any means for engaging with multiple graphs or RDF datasets such as combining graphs and \texttt{FROM} in SPARQL queries.} feature, these mechanisms rely on referencing graphs by an IRI, which is arguably similar to named graphs in RDF datasets. But, as above, the provenance of triples is lost when the union for the shapes graph is computed.
    
    \item When SPARQL is used for, e.g., SPARQL-based constraints, the SPARQL queries are evaluated over an RDF dataset since SPARQL’s evaluation model works on an RDF dataset. In SHACL, the data graph becomes the default graph, and the shapes graph \emph{may} be made available as a named graph whose IRI is accessible via the \texttt{\$shapesGraph} variable. However, SHACL does not specify how this dataset should be constructed when the input itself is an RDF dataset. Moreover, while SHACL explicitly prohibits certain SPARQL features, such as MINUS and SERVICE within SPARQL-based constraints, the dataset-level keywords \texttt{FROM}, \texttt{FROM NAMED}, and \texttt{GRAPH} are allowed, though only applicable to \texttt{\$shapesGraph} as it is the only named graph mentioned in the specification. SHACL discourages the use of \texttt{\$shapesGraph}, however, as: \textit{"[i]t may result in constraints that are not interoperable across different SHACL-SPARQL processors and that may not run on remote RDF datasets."} \cite{shacl} The use of RDF datasets in SPARQL-based constraints has little sense and may lead to interoperability issues. So, one could ask why not reconsider this for RDF dataset validation?
\end{itemize}

While RDF datasets are "flattened" into an RDF graph for SHACL-Core, the use of SHACL-SPARQL reintroduces the notion of RDF datasets created for validation purposes in a limited capacity. However, one could easily conceive how SHACL-SPARQL can be useful for declaring complex rules across graphs if access to the dataset was permitted. 


\subsection{SHACL Implementations}
\label{SHACL Implementations}
We will now examine how existing SHACL implementations handle the validation of RDF datasets. The implementations were selected from the \textit{SHACL Test Suite and Implementation Report} \cite{shacl_test}, which provides an overview of SHACL processors and their conformance to the specification. Most processors can parse RDF serialization formats that support datasets (e.g., TriG \cite{Carothers:14:RT}, N-Quads \cite{Carothers:14:RN}). Still, their handling of RDF datasets differs depending on whether the input represents the data graph, the shapes graph, or both. Table~\ref{tab:imp-rep} summarizes the observed behavior when a dataset is provided as input for either the data or shapes graph.

\begin{table*}[h]
    \caption{Summary of SHACL Implementations and Their RDF Dataset Support}
    \label{tab:imp-rep}
    \scriptsize
    \centering
    \renewcommand{\arraystretch}{1.5}
    \begin{tabular}{|l|l|l|}
    \hline
    \textbf{Name} & \textbf{Data Dataset Support} & \textbf{Shapes Dataset Support} \\\hline
    Corese \cite{corese_shacl} & Data graph is union of all graphs & Shapes graph is union of all graphs \\\hline
    dotNetRDF \cite{dotnetrdf_shacl} & No RDF dataset allowed & No RDF dataset allowed \\\hline
    pySHACL \cite{pyshacl} & Data graph is union of all graphs & Shapes graph is union of all graphs \\\hline
    RDFUnit \cite{rdfunit} & Data graph is default graph & Shapes graph is default graph \\\hline
    shaclex \cite{shaclex} & Data graph is default graph & Shapes graph is default graph \\\hline
    TopBraid \cite{topbraid_shacl}& Data graph is default graph & No RDF dataset allowed \\\hline
    Netage \cite{netage_shacl} & Closed source and unavailable for testing & Closed source and unavailable for testing\\\hline
    \end{tabular}

\end{table*}

As shown in Table~\ref{tab:imp-rep}, none of the surveyed implementations perform dataset-level validation, which comes as no surprise, as SHACL was not conceived for this task. Instead, they either (i) treat the data graph as the union of all graphs, effectively flattening the dataset, (ii) restrict validation to the default graph only, or (iii) yield an error. The first two behaviors obscure the distinction between named graphs and eliminate provenance information that may be essential to the semantics of constraints.

While these approaches may be suitable for specific use cases where treating all graphs as a whole or focusing solely on the default graph is desired, they fail to support more sophisticated scenarios where constraints must apply selectively to certain graphs or span relationships across them. In such cases, developers who need fine-grained control over RDF dataset validation must implement custom preprocessing pipelines or ad-hoc solutions. This demonstrates a clear gap in the current SHACL ecosystem and motivates the need for an explicit and formalized mechanism that allows users to specify precisely how RDF datasets should be validated.

\subsection{SHACL-X}
SHACL-X \cite{shacl_x}\footnote{It is important to note that the proposal was not named as such. We dubbed it "SHACL-X" based on the proposal's namespace, which ended with "shacl-x\#".}  is an extension of SHACL designed to adapt its use to RDF datasets, thereby extending its applicability beyond single RDF graphs. To the best of our knowledge, there is no implementation.


SHACL-X introduces the concept of a \textit{Shapes Dataset}, a collection of shape graphs that group shapes and provide information on how these shape graphs are applied to an RDF dataset. 

To specify where a shapes graph applies, SHACL-X introduces the concept of a \textit{Target Graph}, which identifies the specific graph(s) within the dataset to which a given shapes graph applies. This mechanism is similar to SHACL’s existing notion of targets, but operates at the RDF dataset level.
These targets are declared through the \texttt{shx:targetGraph} predicate. This property can be repeated such that a set of graphs can be selected as the target of a shapes graph. To avoid repeating this property, four reserved IRIs define a collection of graphs:
\texttt{shx:all}, the shape is applied to each graph of the dataset, both named and default;
\texttt{shx:named}, the shape is applied to each of the named graphs;
\texttt{shx:default}, the shape is applied to the default graph;
\texttt{shx:union}, the shape is applied to the union of all graphs (i.e., one graph combining all the triples from named graphs).

Moreover, the concept of an excluded target graph is introduced via the \texttt{shx:targetGraphExclude} predicate to create a less verbose, more complex target definition. When this predicate excludes a graph, it is not included in the target graph selection process. To achieve a defined set of targets, all inclusions are applied before all exclusions. Note that the target definitions should either be specified in the default graph of the shapes dataset or within the specific shapes graph that defines its own target.

To support dataset-aware validation, SHACL-X also extends the validation report model. In SHACL-X, each validation result in a report includes a triple with the
\texttt{shx:resultGraph} predicate, which indicates the graph in which the validation occurred. This helps to identify precisely where in the RDF dataset the error occurred.

While SHACL-X provides a conceptual basis for extending SHACL to RDF datasets, we argue that the proposal \textbf{remains informal and incomplete}. It defines new vocabulary terms but proposes no mapping of constraints to SPARQL-based constraints nor algorithms that would give it a formal foundation. 
\textbf{Graph targeting is limited} to inclusion and exclusion via \texttt{shx:targetGraph} and \texttt{shx:targetGraphExclude}, with no support for set-theoretic combinations of graphs. The proposal offers \textbf{no clear semantics for SPARQL-based constraints or for dataset-level keywords such as \texttt{GRAPH}}, leaving query behavior underspecified. 
\textbf{No prototype implementation or test suite has been released} to evaluate feasibility. Consequently, SHACL-X serves as a conceptual precursor rather than a specification, motivating our work on SHACL-DS, which formalizes these semantics and demonstrates them through an executable implementation.  SHACL-X introduced several concepts that inspired and informed the design of SHACL-DS. While SHACL-DS builds upon these ideas, it refines and extends them to address limitations that we identified in SHACL-X’s expressivity and semantics. 

\section{SHACL-DS}

\begin{figure}
    \centering
    \includegraphics[width=\linewidth]{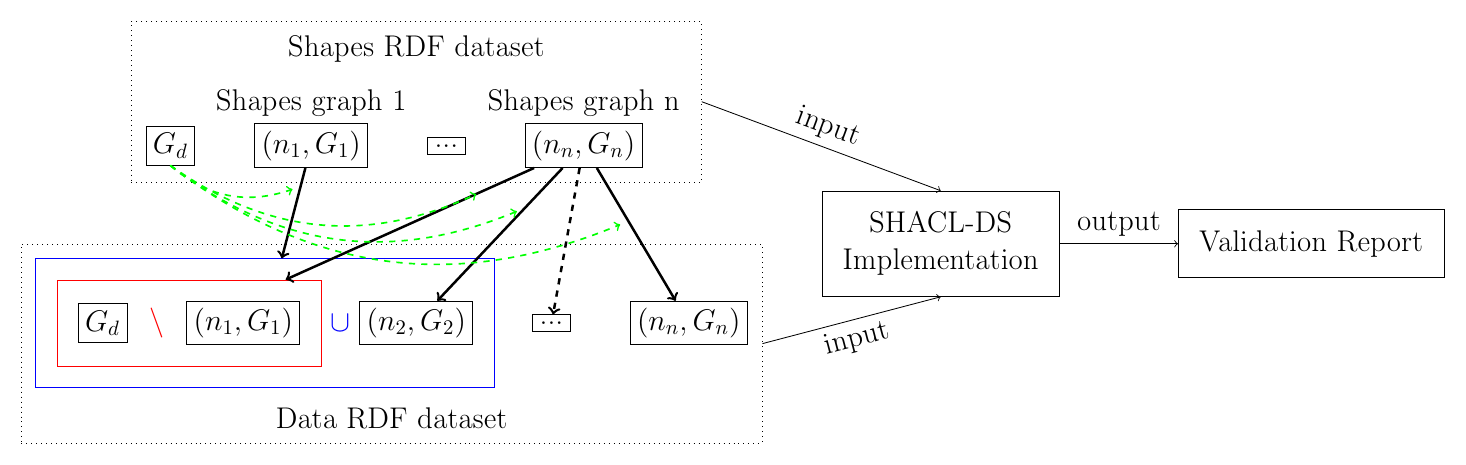}
    \caption{Instead of one shapes graph validating one data graph, SHACL-DS introduces the concept of (named) shapes graphs that can validate one or more (combinations of) (named) data graphs. The red and blue boxes denote graph combinations with their operator. The green arrows denote the relationships between the named graphs in the shapes graph and their targets (themselves shown as black arrows), which are described in the shapes graph's default graph.}
    \label{fig:shaclds}
\end{figure}

SHACL-DS extends\footnote{We emphasize that SHACL-DS extends rather than replaces SHACL. Existing shapes and validators remain applicable.} SHACL with an RDF dataset-level validation model that treats datasets as first-class validation objects. It introduces explicit semantics for (i) declaring which data graphs are validated by which shapes graphs, (ii) forming derived targets through set-theoretic combinations of graphs, (iii) evaluating SPARQL-based constraints in a dataset-aware context, and (iv) reporting validation results with provenance annotations that link each violation to both its data and shape origins. We visually depict (i) and (ii) in Figure \ref{fig:shaclds}.

Let $D=\langle G_d,\{(n_1,G_1), ..., (n_n,G_n)\}\rangle$ be and RDF dataset where $G_d,G_1,..,G_n$ are sets of RDF triples where one, $G_d$, is the default graph and each $(n_i,G_i)$ is a named graph where the graph $G_i$ is identified by the IRI $n_i$. For SHACL-DS, we define the validation of an RDF dataset $D$ as mapping each shapes graph $s_i$ in a shapes dataset $S$ to a set of graphs $\tau_D(s_i)\subseteq D$ that it validates. In the following subsections, we will introduce the SHACL-DS model and, for each aspect of the model, first explain the concepts and vocabulary, followed by the formalism. 

\subsection{Shapes Dataset}

A \textit{Shapes Dataset} is an RDF dataset 
where the named graphs $(n_i,G_i)$ contain SHACL shapes and the default graph $G_d$ relates which (combinations of) graphs in the RDF dataset, called \textit{target graphs}, are validated against what shapes, as illustrated in Figure \ref{fig:shaclds}.\footnote{In this paper, we will often use $s_i$ instead of $(n_i,G_i)$ to refer to a named graph in the shapes dataset.} The figure illustrates that we have access to, among others, set operators to combine graphs. The named graphs of the shapes RDF dataset contain the shapes that triples must respect, and the default graph of the shapes RDF dataset is used to relate the shapes graphs to (combinations of) graphs in the data RDF dataset.

To select these target graphs, SHACL-DS defines two distinct methods: the \textit{Target Graph} definition, which identifies existing graphs in the dataset, and the \textit{Target Graph Combination} definition, which enables selecting a new target graph via set operations on multiple graphs. SHACL-DS also allows each shape's graph to declare its own target independently, but, as examples will show, that is some form of syntactic sugar. When a SHACL shape has no target nodes, then it is omitted from the SHACL validation process. In a similar vein, if a shapes graph has no target graph in SHACL-DS, then it is omitted from the SHACL-DS validation process.

In SHACL, a \textit{focus node} is the RDF node in the data graph to which a shape is applied during validation. Shapes become associated with specific nodes through targeting mechanisms such as \texttt{sh:targetClass}. For each node in a set, all constraints defined in the corresponding shape are evaluated with respect to that node. Similarly, SHACL-DS introduces the concept of a \textit{focus graph}. A focus graph data graph that is validated against a shapes graph. This data graph may be in the dataset or a new graph obtained by applying set operations to multiple graphs.

While we have not yet introduced or formally described the two approaches to target graphs in the \textit{data dataset}, the listing below exemplifies how SHACL shapes are organized in a shapes dataset and how those shapes relate to (named) graphs in the data dataset.

\vspace{0.4cm}

\noindent\includegraphics[width=\linewidth]{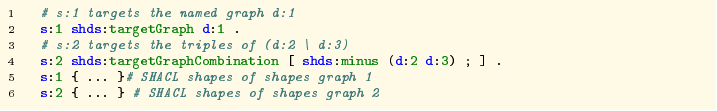}

\subsubsection{Targeting (Named) Graphs}

Target graphs are declared by associating a shapes graph with one or more occurrences of the predicate \texttt{shds:targetGraph}. Each object denotes the IRI of a data graph within the data dataset to which the shapes graph applies. As the default graph is not identified by an IRI, SHACL-DS defines the reserved IRI \texttt{shds:default} to enable the selection of the default graph as a validation target.

To simplify the selection of multiple graphs, SHACL-DS introduces two reserved IRIs: 1) \texttt{shds:named} targets all named graphs in the data dataset. Functionally, this is equivalent to declaring a separate \texttt{shds:targetGraph} triple for each named graph, and ii) \texttt{shds:all} targets every graph in the dataset, including both the default graph and all named graphs.
The predicates and resources \texttt{shds:targetGraph}, \texttt{shds:default}, \texttt{shds:named}, and \texttt{shds:all} were inspired by SHACL-X.

SHACL-DS introduces \texttt{shds:targetGraphExclude}\footnote{\texttt{shds:targetGraphPattern} and \texttt{shds:targetGraphPatternExclude} are also introduced by SHACL-DS behave similarly to shds:targetGraph and shds:targetGraphExclude, but operate on graph patterns rather than fixed IRIs.} for more refined graph selection, which is also adopted from SHACL-X. This allows specific graphs\footnote{The predefined IRIs can also be used in the exclusion.} to be removed from the set of targeted graphs. The final result yields a set of focus graphs, and each graph in that set will be validated against the shapes of the shapes graph.

In the listing below, the \textit{Shapes Dataset} contains a shapes graph that targets all the named graphs. For each of these graphs, Alice must be a Person.

\vspace{0.4cm}

\noindent\includegraphics[width=\linewidth]{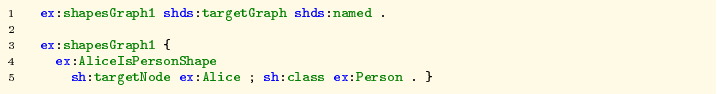}

\subsubsection{Targetting Combinations of Graphs}
The targeting of graph combinations is novel. One motivation for this functionality is the case where data from different sources are stored in distinct named graphs, and one wishes to ensure that a node has at most one value for a property. Thus, in addition to selecting existing graphs as targets, SHACL-DS allows selecting a combination of graphs obtained via set operations. 

\texttt{shds:targetGraphCombination} is used to target such combinations of graphs, and the following operators can be applied to such combinations: i) \texttt{shds:and} to define a target graph as the intersection of 1 or more graphs, ii) \texttt{shds:or} to define a target graph as the union of 1 or more graphs, and \texttt{shds:minus} to define a target graph as the difference between two graphs. The operands of these operators must be either named graphs within the dataset or other graph combinations, allowing nested construction of complex target graphs. The predefined IRIs \texttt{shds:default}, \texttt{shds:named}, and \texttt{shds:all} can also be used within \texttt{shds:or} and \texttt{shds:and} expressions as syntactic sugar. As for \texttt{shds:minus}, however, only \texttt{shds:default} is allowed.

Each \texttt{shds:targetGraphCombination} triple produces a single focus graph during the validation. When formalizing our approach, such triples thus always yield singletons wheresas \texttt{shds:targetGraph} may yield a set.

In the following example, the \textit{Shapes Dataset} contains a shapes graph that targets the union of all graphs minus the \texttt{ex:dataGraph1} named graph. In the focus graph obtained via this combination, every person must know someone.

\vspace{0.2cm}

\noindent\includegraphics[width=\linewidth]{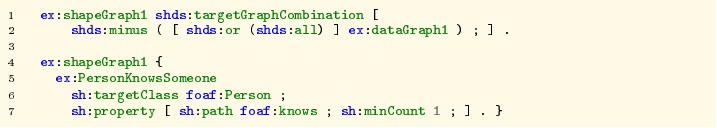}

\vspace{0.2cm}

If a shapes graph declares both explicit targets and target combinations, it is important to note that the excluded graphs only apply to graphs selected via \texttt{shds:targetGraph}. This is intentional: a target graph defined through \texttt{shds:targetGraphCombination} is treated as a newly derived graph and therefore is not subject to exclusion.

\subsubsection{Formalizing Targets}

We have introduced two approaches for targeting graphs or graph combinations. In this section, we formalize how to obtain a set of focus graphs for a given shape graph. Since we allow the construction of graphs using set operators, we must define the set of all graphs that can be constructed from an RDF dataset.

Let $\mathcal{G}(D)$ be the set of all graphs in an RDF dataset $D$. We define the \textit{graph closure} $\mathcal{G}^+(D)$ as the set such that $\mathcal{G}(D) \subseteq \mathcal{G}^+(D)$, and, if $G_1,G_2\in \mathcal{G}^+(D)$, then: $G_1\cup G_2\in \mathcal{G}^+(D)$, $G_1\cap G_2\in \mathcal{G}^+(D)$, and $G_1\setminus G_2\in \mathcal{G}^+(D)$.
All named graphs retain their name, and the default graph can be accessed with \texttt{shds:default}. All constructed graphs have no identifier, but new identifiers (blank node or IRIs) are generated based on the set operations.

Let $D$ be the data dataset and $S$ the shapes dataset. We define the target function $\tau_D=\mathcal{G}(S) \to \mathcal{G}^+(D)$ which maps each shapes graph $s \in \mathcal{G}(S)$ to the set of focus graphs it validates: $\tau_D(s)=T_D(s) \cup C_D(s)$ where both $T_D$ and $C_D$ are functions yielding, respectively, targeted and combined graphs from the data dataset $D$.

Let $T_D^{in}(s)=\{g| \langle s,  \textsf{shds:targetGraph}, g\rangle \wedge g \in \mathcal{G}(D) \}$. Additionally, predefined IRIs expand as follows: if $g=\textsf{shds:default}$, then $g$ is equal to the data dataset's $G_d$; if $g=\textsf{shds:named}$ then, $g$ ranges over all named graphs of the data dataset; if $g=\textsf{shds:all}$, then $g$ ranges over all graphs of the data dataset. Exclusions are defined analogously: $T_D^{out}(s)=\{g|\langle s,  \textsf{shds:targetGraphExclude}, g\rangle \wedge g \in \mathcal{G}(D) \}$
The direct target set is then $T_D(s)=T_D^{in}(s)\setminus T_D^{out}(s)$, which corresponds to the notion of inclusions applied before exclusions.

Each triple $\langle s, \textsf{shds:targetGraphCombination}, C \rangle$ in the RDF description of shapes graph $s$ defines a derived graph through an expression tree built from: \texttt{shds:and}, which is the set intersection of one or more graphs or graph combinations; \texttt{shds:or}, which is the set union of one or more graphs or graph combinations; and \texttt{shds:minus}, which is the set difference of two graphs or graph combinations. The interpretation function $[C]_D \in \mathcal{G}^+(D)$ produces exactly one combined graph. We therefore define $C_D(s)=\{g|\langle s,  \textsf{shds:targetGraphCombination}, C \rangle \wedge [C]_D \in \mathcal{G}^+(D) \wedge g \leftarrow [C]_D\}$. Note that $C_D(s)$ is a set of one or more derived graphs.

\subsection{SPARQL-based Constraints in SHACL-DS}
The $\tau_D$ allows us to determine which (derived) graphs are to be validated by a given shapes graph $s$. Once this selection process is completed, the SHACL-DS validation process is similar to SHACL: validating a (derived) focus graph $d \in \mathcal{G}^+(D)$ against a shapes graph $s \in \mathcal{G}(S)$ is the same as validating a single data graph against a shapes graph in SHACL that only uses SHACL core constraints.

As seen in Section \ref{SHACL Implementations}, not only does SHACL use an "internal" RDF Dataset for SPARQL-based constraints, but the behavior when giving RDF datasets as input differs across implementations, as such behavior has been left to implementations. Moreover, SHACL does not prohibit the use of dataset-level SPARQL keywords such as NAMED, FROM NAMED, and GRAPH, leaving their intended behavior ambiguous. In this section, we address this problem by specifying how FROM, FROM NAMED, and GRAPH clauses in SPARQL-based constraints are to be interpreted and how they interplay with the (derived) focus graphs in SHACL-DS.

\subsubsection{Validation Context}
SPARQL queries must be evaluated against the RDF dataset that is the data dataset. In other words, the data dataset must not be "flattened." As SHACL provides no formal semantics for which graphs are visible to a SPARQL-based constraint, a solution must be found. SHACL-DS resolves this by defining an RDF dataset-aware evaluation model for SPARQL-based constraints.

We note that, similar to SHACL, we do not define the behavior of the pre-bound variables \texttt{\$shapesGraph} and \texttt{\$current\allowbreak Shape} in a SPARQL-based constraint. SHACL states that these variables are optional and should be avoided as they are not guaranteed to be interoperable across SHACL implementations. As SHACL-DS extends SHACL and strives for backward compatibility, it maintains this stance.

\subsubsection{RDF Dataset Views}
\begin{figure*}
    \centering
    \includegraphics[width=\linewidth]{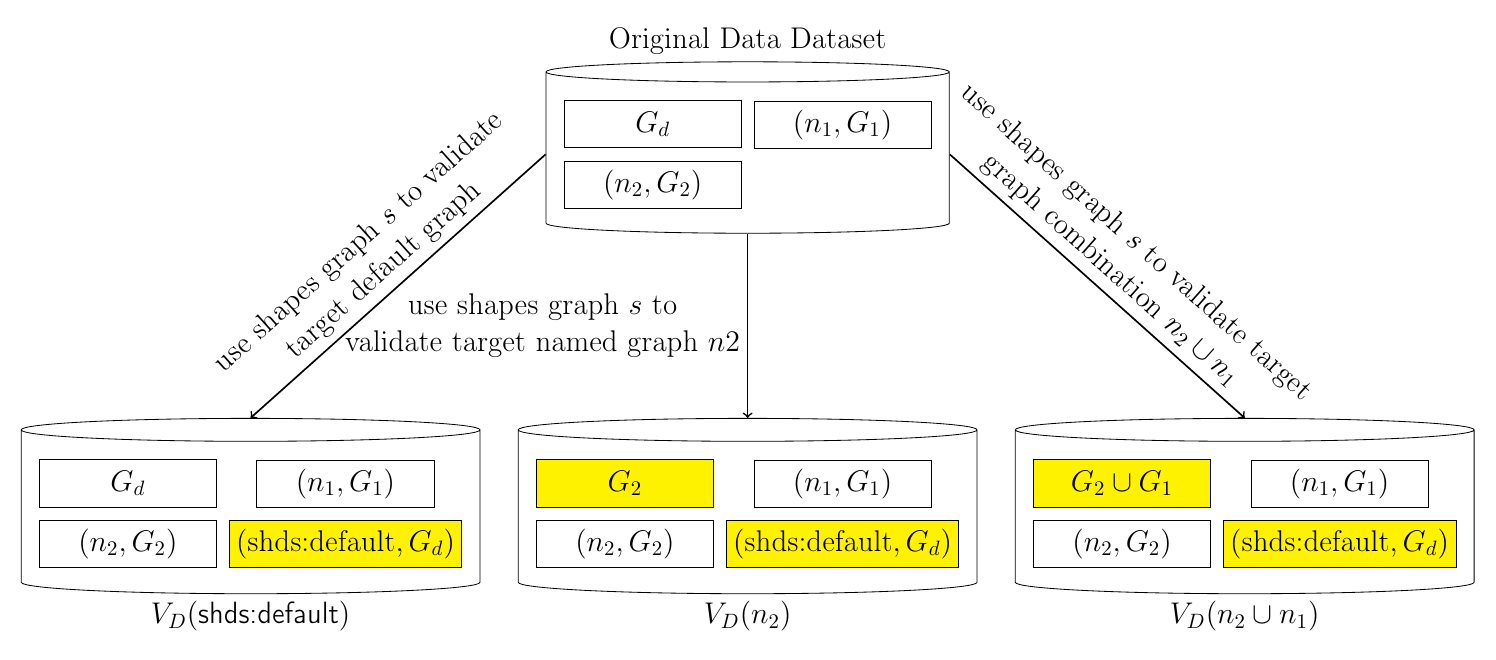}
    \caption{Examples of constructing dataset views based on targets.}
    \label{fig:example-views}
\end{figure*}

We enable the aforementioned RDF-dataset awareness by introducing the concept of a \textit{Dataset View}.
For each shapes graph $s \in \mathcal{G}(S)$ and each focus graph $g \in \tau_D(s)$, SHACL-DS constructs an \textit{evaluation dataset} $V_D(g)$. This RDF dataset is based on the original data dataset $D$, but the original default graph $G_d$ is stored in a named graph with IRI \texttt{shds:default} and the focus graph becomes the default graph, as exemplified in Figure \ref{fig:example-views}.

With this approach, all graphs in $\mathcal{G}(D)$, thus also including the default graph, remain available as named graphs. By making the original default graph available in the named graph \texttt{shds:default}, one can write SPARQL-based queries that can reliably rely on that original default graph. I.e., we preserve access to all the data of the original data dataset. This enables the same SPARQL-based constraint to operate consistently across different graphs or graph combinations.

All SHACL-core constraints are tested against an evaluation dataset's default graph. But, since SHACL-DS defines an evaluation dataset, keywords such as \texttt{GRAPH}, \texttt{FROM}, and \texttt{FROM NAMED} now make sense and are permitted when SPARQL queries are used in SHACL-DS. We continue to prohibit SPARQL features forbidden by SHACL-SPARQL, ensuring compatibility with the SHACL specification.

By specifying how SPARQL queries with dataset-level keywords should be evaluated, SHACL-DS improves SHACL’s ability to validate RDF datasets.
For example, given the following RDF data dataset.

\noindent\includegraphics[width=\linewidth]{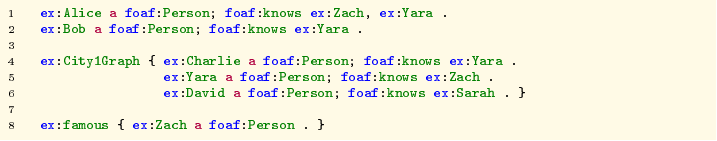}

\vspace{0.2cm}

In the listing below, the \textit{Shapes Dataset} contains a shapes graph that targets all the graphs except the \texttt{ex:famous}. For each of these graphs, every person must (indirectly) know at least one famous person (i.e., connected to a famous person via one or more \texttt{foaf:knows} predicates). 
In this example, the default graph yields an error: \texttt{ex:Bob} is not (indirectly) connected to a famous person. In the \texttt{ex:City1Graph} graph, \texttt{ex:David} does not satisfy the constraint for the same reason. 

\vspace{0.2cm}

\noindent\includegraphics[width=\linewidth]{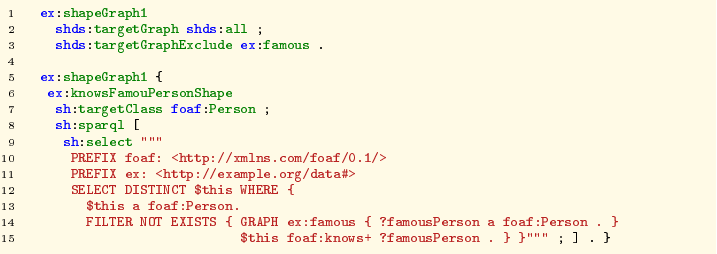}

\vspace{0.2cm}

Below, we present another example using target combinations of graphs. In this example, we validate the union of the default graph and \texttt{ex:City1Graph} using the same SPARQL-based constraint. When the two graphs are combined, only one person yields an error, namely \texttt{ex:David}, as \texttt{ex:Bob} is now indirectly connected to a famous person via \texttt{ex:Yara}.

\vspace{0.4cm}
\noindent\includegraphics[width=\linewidth]{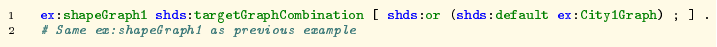}

\subsection{Validation Reports}

To enhance the validation reporting process in SHACL-DS, we introduce two new predicates: \texttt{shds:sourceShapeGraph} and \texttt{shds:focusGraph}. These predicates provide additional context to validation results by annotating them with information about the originating shapes graph and the relevant data graph. Essentially, these predicates are the dataset-level equivalent of \texttt{sh:sourceShape} and \texttt{sh:focusNode}.

We use \texttt{shds:sourceShapeGraph} to identify the specific shapes graph within the shapes dataset that contained the constraint responsible for generating the validation result. This allows users to trace validation errors back to their corresponding shape definitions, disambiguating shapes sharing the same IRI in different shapes graphs.

We use \texttt{shds:focusGraph} to indicate the focus graph, a specific data graph within the data dataset where the focus node originated. If the validated node belongs to a named graph, the focus data graph is the IRI of that named graph. If the validated node originates from the default graph, \texttt{shds:focusGraph} is assigned the reserved IRI \texttt{shds:default}. When the focus graph is a combination of multiple graphs, it does not correspond to a graph in the dataset\footnote{Excluding the particular case where the combination is simply a named graph or \texttt{shds:default}}, a copy of the explicit combination declaration should be included in the validation report. This structure ensures that the validation result accurately identifies the (combined) target data graph. Given the verbosity of such reports, we refer the reader to the test cases in the repository for examples.

\vspace{0.3cm}
\begin{algorithm}[H]
\SetAlgoLined
\KwIn{$S$ an RDF shapes dataset}
\KwIn{$D$ an RDF data dataset}

$vr \leftarrow \{\}$; // initialize validation results 

\ForEach{Shapes graph $s \in S$}{
    \ForEach{Target graph $t \in \tau_D(s)$}{
        // add results of validating the evaluation dataset against s
        
        $vr \leftarrow vr \cup validate(V_D(t), s)$
    }
}
\Return{createValidationReportGraph$(vr)$}\;
\caption{RDF Dataset Validation with SHACL-DS}
\label{alg:1}
\end{algorithm}

\section{Prototype Implementation}
As a proof of concept, a prototype SHACL-DS engine was implemented. This prototype extends the SHACL module of dotNetRDF (see Section \ref{SHACL Implementations}). It introduces support for dataset-level validation while maintaining compatibility with existing SHACL functionality. The implementation follows Algorithm \ref{alg:1} and the evaluation dataset views are constructed in memory. Once validation is performed, the validation results are extended with SHACL-DS validation annotations, namely \texttt{shds:focusGraph} and \texttt{shds:sourceShapeGraph}. These annotated reports are then aggregated into a single validation report. For SPARQL-based constraints, the SHACL-DS engine applies queries directly to the Data Dataset.

\subsection{Test Cases}

The repository includes a test suite developed to guide developers in creating their SHACL-DS implementations and to provide examples. These test cases test SHACL-DS's functionality, covering both the introduced features and complex validation scenarios.
The test cases cover: 1) Simple target graph declarations using a named graph or the predefined IRIS; 2) Proper exclusion of a graph from the set of target graphs; 3) Simple target graph combination declarations using each set operator on two graphs; 4) Complex target graph combination declarations with nested graph combinations, which may include predefined IRIs; and 5) Constraints using SPARQL queries with dataset-level keywords.

\subsection{Validating SHACL-DS with a Shapes Dataset}
SHACL provides a shapes graph for validating SHACL shapes, enabling one to ensure they are "well-formed." Similarly, SHACL-DS provides a Shapes Dataset\footnote{Available at \url{https://w3id.org/shacl-ds/shacl-ds#}} to validate a Shapes Dataset. This ensures that target graph declarations using the \texttt{shds:targetGraph} and \texttt{shds:targetGraphCombination} predicates are well defined. For Target Graph Declarations, it ensures that targets are IRIs. These are one of the predefined IRIs (\texttt{shds:default}, \texttt{shds:named}, \texttt{shds:all}) or should be the IRI of a named graph in the Data dataset. For Target Graph Combination Declarations, it ensures that targets are graph combinations. These are either the IRI of a named graph or a blank node with only one of the properties \texttt{shds:or}, \texttt{shds:and} or \texttt{shds:minus}. It also ensures that the objects of these properties are well-defined SHACL Lists of graph combinations. This list must also be of length two for \texttt{shds:minus}.

\section{Discussion}
\subsection{SHACL-DS vs SHACL-X}
\label{shaclx_vs_shaclds}
While SHACL-X initially inspired SHACL-DS, we refined some ideas, such as more flexible graph-targeting for validation and the ability to validate an RDF dataset with SPARQL-based constraints. Where we "limited" flexibility compared to SHACL-X is the subject of targets; where SHACL-X allows the range of targets to include individual shapes, named graphs containing shapes, or the entire shapes dataset, SHACL-DS restricts the range of targets to named graphs only. We have chosen a simpler approach, which also simplified the formalism and implementation. 

SHACL-X also introduced \texttt{shx:include} to share shapes between named graphs. While these features provide additional flexibility, SHACL-DS does not currently include them, as it focuses on foundational concepts for RDF dataset validation. Future work could look into the inclusion of such functionality. While overall, SHACL-DS builds upon concepts introduced by SHACL-X, it refines them to ensure practical applicability and implementation feasibility. By integrating these refinements and providing a prototype, SHACL-DS takes a concrete step toward enabling dataset-level validation within SHACL.

\subsection{Prototope Limitations}
The implementation is a prototype that focuses on correctness rather than performance optimization. No benchmarking has been conducted to assess its efficiency on large-scale RDF datasets. The implementation furthermore relies on an existing SHACL codebase that not only loads the datasets into main memory but also the evaluation datasets. Future work could look into 1) a bespoke implementation for SHACL-DS, and 2) investigate means to optimize the validation of such datasets. For example, our algorithm in Algorithm \ref{alg:1} creates, for each shape's graph and each of its targets, an evaluation dataset. One could consider applying shape graphs to shared targets or prioritizing combined graph combinations when they are reused as part of a more complex graph.

\section{Conclusion}
While SHACL is an established technique for RDF graph validation, there is a need for RDF dataset validation. Evidence for this stems from our project and SHACL-X, a proposal for RDF dataset validation. SHACL-X has some limitations, but, more importantly, no implementation.

We introduced SHACL-DS, an approach that builds upon SHACL to enable native validation of RDF datasets. SHACL-DS introduces 1) concepts such as Shapes Datasets, Target Graph and Target Combination Declarations, and a vocabulary to represent those, 2) prescribes how SPARQL-based constraints should be evaluated against (combined) target graphs, 3) formalizes the RDF dataset validation process, and 4) implements the process. The whole provides a structured mechanism for expressing validation constraints across graphs while preserving information about the dataset’s structure. Our implementation, built as an extension to the dotNetRDF, demonstrates the feasibility of SHACL-DS in practice. We developed a structured test suite covering fundamental and complex scenarios, allowing others to build similar engines. The tests also serve as a demonstrator. These contributions lead to the lesson we can learn from this paper: RDF dataset validation is possible by building on SHACL. 

For future work, we aim to assess the applicability of SHACL-DS and the performance of the implementation through use cases, beyond the use case thatinformed this study, and community feedback. As for the functionalities, other aspects that we wish to explore are additional features, such as \texttt{shx:include} proposed by SHACL-X. While such extensions are nice to have, we believe, and therefore prioritize approaches to optimize RDF Dataset validation.
\newpage
\appendix
\section{Namespaces}
All listings contain fragments of RDF datasets in Trig \cite{Carothers:14:RT} that use the following namespace prefix bindings: 

{\small
\begin{itemize}
\item \texttt{sh:}  \texttt{http://www.w3.org/ns/shacl\#}
\item \texttt{shx:} \texttt{http://www.w3.org/ns/shacl-x\#}
\item \texttt{shds:} \texttt{http://www.w3id.org/shacl-ds\#}
\item \texttt{foaf:} \texttt{http://xmlns.com/foaf/0.1/}
\item \texttt{ex:} \texttt{http://example.org/}
\item \texttt{s: http://example.org/shapes/}
\item \texttt{d: http://example.org/data/}
\end{itemize}
}

\paragraph*{Supplemental Material Statement:} 
Source code, documentation, and test-cases are available at the following  GitHub repository or Zenodo archive:
\begin{itemize}
    \item Github: \url{https://github.com/Ikeragnell/SHACL-DS}
    \item Zenodo: \url{https://doi.org/10.5281/zenodo.18771029}
\end{itemize}
\paragraph{Use of Generative AI:}
We used Grammarly and ChatGPT to assist with writing and proofreading. ChatGPT was also used to condense and rephrase sections to meet the page limit and to use and debug LaTeX TikZ for some figures. We also used Alternative Text Assistant to assist with the figures alternative text. All scientific content, results, and conclusions were developed solely by the authors.


\bibliographystyle{splncs04}
\bibliography{references}


\end{document}